# Wireless Power Transfer in Cooperative DF Relaying Networks with Log-Normal Fading

Khaled M. Rabie*, Bamidele Adebisi* and Mohamed-Slim Alouini‡

*School of Electrical Engineering, Manchester Metropolitan University, Manchester, UK, M15 6BH

‡King Abdullah University of Science and Technology (KAUST), Thuwal, Mekkah Province, Saudi Arabia

Email: {k.rabie@mmu.ac.uk; b.adebisi@mmu.ac.uk; slim.alouini@kaust.edu.sa}

*Abstract*—Energy-harvesting and wireless power transfer in cooperative relaying networks have recently attracted a considerable amount of research attention. Most of the existing work on this topic however focuses on Rayleigh fading channels which represents outdoor environments. Unlike these studies, in this paper we analyze the performance of wireless power transfer in two-hop decode-and-forward (DF) cooperative relaying systems in indoor channels characterized by log-normal fading. Three well-known energy-harvesting protocols are considered in our evaluations: a) time-switching relaying (TSR), b) power-splitting relaying (PSR) and c) ideal relaying receiver (IRR). The performance is evaluated in terms of the ergodic outage probability for which we derive accurate analytical expressions for the three systems under consideration. Results show that careful selection of the energy-harvesting time and power-splitting factors in the TSR- and PSR-based system are important to optimize performance. It is also presented that the optimized PSR system has near-ideal performance and that increasing the source transmit power and/or the energy-harvester efficiency can further improve performance.

## I. INTRODUCTION

THE capability of electromagnetic waves of concurrently carrying information and energy signals, an approach also known as simultaneous wireless information and power transfer (SWIPT), has recently attracted considerable research interest. Although many studies have analyzed the performance of point-to-point SWIPT based systems [1]–[3], cooperative relaying SWIPT networks, where an intermediate relaying node is used to forward the source's data to the intended destination, have been by far more extensively investigated in the literature, see e.g. [4]–[7] and the reference therein. More specifically, the authors in [5] examined the performance of a half-duplex amplify-and-forward (AF) relaying network with energy-harvesting where a greedy switching policy, i.e. the relaying node only transmits when its residual energy grantees successful decoding at the destination, was investigated. Later on, the authors of [6] evaluated the performance of a one-way AF relaying system with three different energy-harvesting relaying protocols, namely, time-switching relaying (TSR), power-splitting relaying (PSR) and ideal relaying receiver (IRR). Furthermore, [8] considered the outage probability and ergodic capacity analysis of a two-way energy-harvesting relaying network. Decode-and-forward (DF) relaying with energy-harvesting was studied in [9], [10]. In addition, the performance of energy-constrained multiple-relay networks with relay selection is examined in [11].

All the aforementioned work however is limited to Rayleigh fading which is used to model the outdoor wireless channel. Only recently, the authors of [12] have analyzed the performance of a dual-hop AF-based SWIPT system over log-normal fading channels. In contrast, in this paper, we study the performance of a dual-hop SWIPT system with DF relaying over the log-normal channel in terms of the ergodic outage probability. It is important to stress here that the ergodic outage probability analysis of the proposed DF system is fundamentally different from that of AF-based approach previously investigated in [12], due the distinct nature of each relaying protocol. Three well-known energy-harvesting protocols are studied in this work, namely, TSR, PSR and IRR.

Therefore, the contribution of this paper is as follows. First, we derive analytical expressions for the ergodic outage probability for the TSR, PSR and IRR-based systems over log-normal fading channels. We then address the optimization problem of the energy-harvesting time and power-splitting factor of the TSR and PSR approaches. Results reveal that optimizing the energy-harvesting and power-splitting factors will minimize the ergodic outage probability. In addition, it is shown that increasing the source transmit power and/or the energy-harvester efficiency can further improve the system performance.

The rest of this paper is organized as follows. Section II describes the system model. Sections III, IV and V are dedicated to analyze the ergodic outage probability of the TSR, PSR and IRR approaches, respectively. Numerical examples and simulation results are presented and discussed in Section VI. Finally, Section VII concludes the paper and outlines the main results.

## II. SYSTEM MODEL

Fig. 1 illustrates the system under consideration which consists of a single-antenna source node, relay node and destination node. In this system, the source node transmits its data, with power $P_s$, to the destination via an energy-constrained DF relay. It is assumed that there is no direct link between the source and destination nodes; hence the end-to-end communication is accomplished over two phases. We also assumed that the relay has no external power supply, i.e. it is entirely dependent on harvesting the energy signal transmitted by the source, and that the power consumed by the circuitry to process data at the relay is neglected.

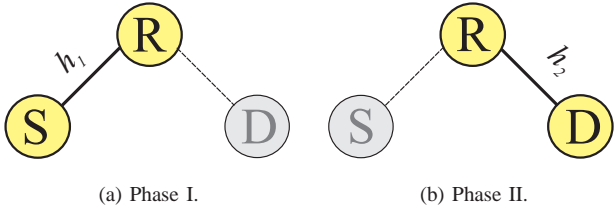

(a) Phase I.          (b) Phase II.

Figure 1: System model.

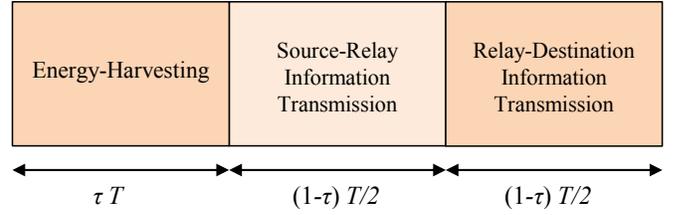

Figure 2: Time frame structure in the TSR protocol.

The source-relay and relay-destination channel coefficients are denoted by $h_1$ and $h_2$ with $d_1$ and $d_2$ being the corresponding distances, respectively. The channels remain constant over the block time, $T$, and vary independently and identically from one block to another according to log-normal distribution with the following probability density function (PDF)

$$f_Z(z_i) = \frac{\xi}{\sqrt{2\pi}\sigma z_i} \exp\left[-\frac{(10\log_{10}(z_i) - \mu)^2}{2\sigma^2}\right], \quad (1)$$

where $z_i = h_i^2$, $i \epsilon \{1, 2\}$, $\xi = 10/\ln(10)$ is a scaling constant, $\mu$ and $\sigma^2$ (both in decibels) are the mean and the standard deviation of $10 \log_{10}(h)$, respectively. As mentioned in the introduction, the system performance is evaluated in terms of the ergodic outage probability. This probability is defined as the probability that the instantaneous capacity falls below a certain threshold value ($C_{th}$) and can be calculated for the proposed DF relaying system as

$$\mathcal{O}(C_{th}) = \Pr\{\min\{C_r(\gamma_r), C_d(\gamma_d)\} < C_{th}\}, \quad (2)$$

where $\min\{.\}$ represents the minimum argument, $C_r$ and $C_d$ are the instantaneous capacities of the source-relay and relay-destination links, respectively, and $\gamma_r$ and $\gamma_d$ denote the corresponding signal-to-noise ratios (SNRs) at the relay and destination nodes.

## III. Ergodic Outage Probability Analysis of the Time-Switching Relaying System

In this protocol, the time required to transmit one block from the source to the destination, also referred as the time frame, given by $T$, is divided into three time slots as shown in Fig. 2. The first time period is the energy-harvesting time, $\tau T$, during which the relay harvests the power signal broadcast by the source node, where $0 \leq \tau \leq 1$ is the energy-harvesting time factor. The remaining time is divided into two time slots each of length $(1 - \tau) T/2$, used for data transmission during phase I (source-relay transmission) and phase II (relay-destination transmission).

To begin with, the received signal at the relay in the first phase can be expressed as

$$y_r(t) = \sqrt{\frac{P_s}{d_1^m}} h_1 s(t) + n_r(t), \quad (3)$$

where $s(t)$ is the information signal normalized as $\mathbb{E}\left[|s|^2\right] = 1$, $m$ is the path loss exponent and $n_r$ is the additive white Gaussian noise (AWGN) signal at the relay node with variance $\sigma_r^2$. Therefore, the harvested energy at the relay can be written as

$$E_H = \eta \, \tau T \, \frac{P_s h_1^2}{d_1^m}, \quad (4)$$

where $0 < \eta < 1$ is the energy-harvesting efficiency determined mainly by the circuitry. Now, the received signal at the destination node can be expressed as

$$y_d(t) = \sqrt{\frac{P_r}{d_2^m}} h_2 \, \bar{s}(t) + n_d(t), \quad (5)$$

where $\bar{s}(t)$ is the decoded version of the source signal, $n_d(t)$ is the noise at the destination node with variance $\sigma_d^2$ and $P_r$ is the relay transmit power which is related to the harvested energy as

$$P_r = \frac{E_H}{(1-\tau) T/2} = \frac{2 \eta P_s h_1^2}{d_1^m} \frac{\tau}{(1-\tau)}. \quad (6)$$

Substituting (6) into (5) yields

$$y_d(t) = \sqrt{\frac{2 \eta \tau P_s}{(1-\tau) d_1^m d_2^m}} h_1 h_2 \, \bar{s}(t) + n_d(t). \quad (7)$$

Grouping the information and noise terms in (3) and (7), we can obtain the SNRs at the relay and destination nodes, respectively, as follows

$$\gamma_r = \frac{P_s \, h_1^2}{d_1^m \sigma_r^2} \quad (8)$$

and

$$\gamma_d = \frac{2 \eta \tau P_s \, h_1^2 \, h_2^2}{(1-\tau) \, d_1^m \, d_2^m \sigma_d^2}. \quad (9)$$

Since in the TSR protocol information transmission takes place only during the time fraction $(1 - \tau)$, the instantaneous capacity of the source-relay and relay-destination links of this system is given by

$$C_i^{TSR} = \frac{(1-\tau)}{2} \log_2(1 + \gamma_i) \quad (10)$$

where $i \epsilon \{r, d\}$.

To derive the ergodic outage probability for this system, we first write (2) as

$$\begin{aligned}
\mathcal{O}^{TSR}(C_{th}) &= \Pr\left\{\min\left\{C_r^{TSR}, C_d^{TSR}\right\} < C_{th}\right\} \\
&= 1 - \Pr\left\{\min\left\{C_r^{TSR}, C_d^{TSR}\right\} \geq C_{th}\right\} \\
&= 1 - \Pr\left\{C_r^{TSR} \geq C_{th}, C_d^{TSR} \geq C_{th}\right\} \\
&= 1 - \underbrace{\Pr\left\{C_r^{TSR} \geq C_{th}\right\}}_{\mathcal{O}_1^{TSR}(C_{th})} \\
&\quad + \underbrace{\Pr\left\{C_r^{TSR} \geq C_{th}, C_d^{TSR} < C_{th}\right\}}_{\mathcal{O}_2^{TSR}(C_{th})}.
\end{aligned} \tag{11}$$

It is clear that the ergodic outage probability requires calculating two probabilities. Using (8) and (10), and substituting $X = h_1^2$, the first probability in (11) can be obtained as

$$\begin{aligned}
\mathcal{O}_1^{TSR}(C_{th}) &= \Pr\left\{C_r^{TSR} \geq C_{th}\right\} \\
&= \Pr\left\{\frac{(1-\tau)}{2}\log_2\left(1 + \frac{P_s X}{d_1^m \sigma_r^2}\right) \geq C_{th}\right\} \\
&= \Pr\left\{\frac{P_s X}{d_1^m \sigma_r^2} \geq U\right\} \\
&= \Pr\left\{X \geq \Omega U\right\} \\
&= 1 - F_X(\Omega U),
\end{aligned} \tag{12}$$

where $U = 2^{\frac{2C_{th}}{1-\tau}} - 1$, $\Omega = d_1^m \sigma_r^2 / P_s$ and $F_X(.)$ denotes the cumulative distribution function (CDF) of the RV $X$. Since $X$ is log-normally distributed, its CDF can be given by

$$F_X(\Omega U) = \frac{1}{2}\left(1 + \mathrm{erf}\left[\frac{\xi \ln(\Omega U) - 2\mu_{h_2}}{2\sqrt{2}\sigma_{h_2}}\right]\right). \tag{13}$$

Now, using (8)−(10), and substituting $Y = h_2^2$, the second probability in (11) can be calculated as

$$\begin{aligned}
\mathcal{O}_2^{TSR}(C_{th}) &= \Pr\left\{C_r^{TSR} \geq C_{th}, C_d^{TSR} < C_{th}\right\} \\
&= \Pr\left\{X \geq \Omega U, \frac{2\eta\tau P_s XY}{(1-\tau)d_1^m d_2^m \sigma_d^2} < U\right\} \\
&= \Pr\left\{X \geq \Omega R, Y < \frac{\Lambda U}{X}\right\}
\end{aligned} \tag{14}$$

where $\Lambda = (1-\tau)d_1^m d_2^m \sigma_d^2 / 2\eta\tau P_s$.

Using the PDF and CDF of the log-normally distributed RVs $X$ and $Y$, we can calculate the probability in (14) as

$$\mathcal{O}_2^{TSR}(C_{th}) = \int_{\Omega U}^{\infty} f_X(z) F_Y\left(\frac{\Lambda U}{z}\right) \mathrm{d}z, \tag{15}$$

where

$$f_X(z) = \frac{\xi}{z\sqrt{8\pi\sigma_{h_1}^2}}\exp\left[-\frac{(\xi\ln(z) - 2\mu_{h_1})^2}{8\sigma_{h_1}^2}\right] \tag{16}$$

and

$$F_Y\left(\frac{\Lambda U}{z}\right) = \frac{1}{2}\left(1 + \mathrm{erf}\left[\frac{\xi\ln\left(\frac{\Lambda U}{z}\right) - 2\mu_{h_2}}{2\sqrt{2}\sigma_{h_2}}\right]\right), \tag{17}$$

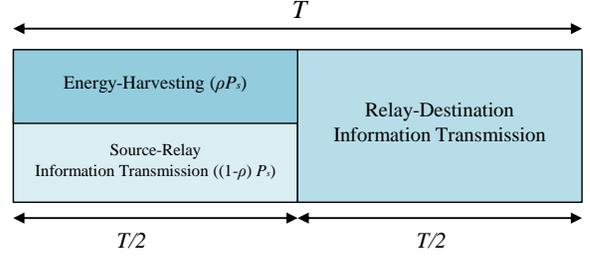

Figure 3: Time frame structure of the PSR protocol.

where $\mathrm{erf}[.]$ is the error function given by

$$\mathrm{erf}(x) = \frac{2}{\sqrt{\pi}}\int_0^x \exp\left(-t^2\right)\mathrm{d}t. \tag{18}$$

Finally, substituting (12) and (15) into (11) yields the system ergodic outage probability, given by (19), shown at the top of the next page.

## IV. Ergodic Outage Probability Analysis of the Power-Splitting Relaying System

In this protocol, the block time, $T$, is divided evenly for the source-relay and relay-destination transmissions as illustrated in Fig. 3. During the first half, the relay allocates a portion of the received signal power, $\rho P$, to the energy-harvester whereas the remaining power, $(1 - \rho)P$, is used for information transmission, where $0 \leq \rho \leq 1$ is the power-splitting factor. Therefore, in the first phase, the received signal at the input of the energy-harvester can be expressed as

$$\sqrt{\rho}y_r(t) = \sqrt{\frac{\rho P_s}{d_1^m}}h_1 s(t) + \sqrt{\rho}\,n_{a,r}(t), \tag{20}$$

where $n_{a,r}(t)$ is the relay antenna noise with variance $\sigma_{a,r}^2$.

Using (20), the harvested energy at the relay node in the first phase can be simply written as

$$E_H = \frac{\eta\rho P_s h_1^2 T}{2d_1^m}. \tag{21}$$

On the other hand, the base-band signal at the information receiver, $\sqrt{1-\rho}y_r(t)$, can be given by

$$\begin{aligned}
\sqrt{1-\rho}\,y_r(t) &= \sqrt{\frac{(1-\rho)P_s}{d_1^m}}h_2 s(t) \\
&\quad + \sqrt{1-\rho}\,n_{a,r}(t) + n_{c,r}(t),
\end{aligned} \tag{22}$$

where $n_{c,r}(t)$ is the noise signal at the information receiver with variance $\sigma_{c,r}^2$.

In the second phase, the DF relay decodes the signal in (22), re-modulates and forwards it using the harvested energy in (21). Therefore, the received signal at the destination node in the second phase can be expressed as

$$y_d(t) = \sqrt{\frac{P_r}{d_2^m}}h_2 \bar{s}(t) + n_d(t), \tag{23}$$

$$\mathcal{O}_{TSR}\left(C_{th}\right) = 1 - \frac{1}{2}\mathrm{erfc}\left[\frac{\xi\ln\left(\Omega U\right) - 2\mu_{h_2}}{2\sqrt{2}\sigma_{h_2}}\right] + \frac{\xi}{\sqrt{32\pi\sigma_{h_1}^2}}\int\limits_{\Omega U}^{\infty}\frac{1}{z}\exp\left[-\frac{\left(\xi\ln\left(z\right) - 2\mu_{h_1}\right)^2}{8\sigma_{h_1}^2}\right]\left(1 + \mathrm{erf}\left[\frac{\xi\ln\left(\frac{AU}{z}\right) - 2\mu_{h_2}}{2\sqrt{2}\sigma_{h_2}}\right]\right)\mathrm{d}z. \tag{19}$$

where $P_r$ is the relay transmit power which is related to the harvested energy as

$$P_r = \frac{2E_H}{T/2} = \frac{\eta\rho P_s h_1^2}{d_1^m}. \tag{24}$$

Now, substituting (24) into (23) produces

$$y_d\left(t\right) = \sqrt{\frac{\eta\rho P_s}{d_1^m d_2^m}}\,h_1 h_2\,\bar{s}\left(t\right) + n_d\left(t\right). \tag{25}$$

Using (22) and (25), the SNRs at the relay and destination nodes can be, respectively, expressed as

$$\gamma_r = \frac{\left(1 - \rho\right)P_s h_1^2}{d_1^m \sigma_r^2} \tag{26}$$

and

$$\gamma_d = \frac{\eta\rho P_s\,h_1^2 h_2^2}{d_1^m d_2^m \sigma_d^2}. \tag{27}$$

The instantaneous capacity at the relay and destination, for the PSR based system, is given by

$$C_i^{PSR} = \frac{1}{2}\log_2\left(1 + \gamma_i\right) \tag{28}$$

where $i\epsilon\left\{r, d\right\}$ and the factor $\frac{1}{2}$ is due to the fact that the end-to-end communication is accomplished over two phases.

Similarly as in the TSR system, the ergodic outage probability of the PSR approach can be calculated as

$$\mathcal{O}_{PSR}\left(C_{th}\right) = 1 - \underbrace{\Pr\left\{C_r^{PSR} \geq C_{th}\right\}}_{\mathcal{O}_1^{PSR}\left(C_{th}\right)} + \underbrace{\Pr\left\{C_r^{PSR} \geq C_{th}, C_d^{PSR} < C_{th}\right\}}_{\mathcal{O}_2^{PSR}\left(C_{th}\right)}. \tag{29}$$

Using (26) and (28), the first probability in (29) can be written as

$$\begin{aligned}
\mathcal{O}_1^{PSR}\left(C_{th}\right) &= \Pr\left\{C_r^{PSR} \geq C_{th}\right\} \\
&= \Pr\left\{\frac{1}{2}\log_2\left(1 + \frac{\left(1 - \rho\right)P_s X}{d_1^m \sigma_r^2}\right) \geq C_{th}\right\} \\
&= \Pr\left\{\frac{\left(1 - \rho\right)P_s X}{d_1^m \sigma_r^2} \geq R\right\} \\
&= \Pr\left\{X \geq \Xi R\right\} \\
&= 1 - F_X\left(\Xi R\right),
\end{aligned} \tag{30}$$

where $R = 2^{2C_{th}} - 1$, $\Xi = d_1^m \sigma_r^2 / \left(1 - \rho\right)P_s$ and $F_X\left(.\right)$ represents the CDF of $X$ which can be given in this case as

$$F_Y\left(\Xi R\right) = \frac{1}{2}\left(1 + \mathrm{erf}\left[\frac{\xi\ln\left(\Xi R\right) - 2\mu_{h_2}}{2\sqrt{2}\sigma_{h_2}}\right]\right). \tag{31}$$

Using (26)−(28), the second probability in (29) can be determined as follows

$$\begin{aligned}
\mathcal{O}_2^{PSR}\left(C_{th}\right) &= \Pr\left\{C_r^{PSR} \geq C_{th}, C_d^{PSR} < C_{th}\right\} \\
&= \Pr\left\{X \geq \Xi R, \frac{\eta\rho P_s XY}{d_1^m d_2^m \sigma_d^2} < R\right\} \tag{32} \\
&= \Pr\left\{X \geq \Xi R, Y < \frac{\Upsilon R}{X}\right\}, \tag{33}
\end{aligned}$$

where $\Upsilon = d_1^m d_2^m \sigma_d^2 / \eta\rho P_s$.

Using the PDF and CDF of the RVs $X$ and $Y$, we can calculate the probability in (33) as

$$\mathcal{O}_2^{PSR}\left(C_{th}\right) = \int\limits_{\Xi R}^{\infty} f_X\left(z\right) F_Y\left(\frac{\Upsilon R}{z}\right)\mathrm{d}z, \tag{34}$$

where

$$f_X\left(z\right) = \frac{\xi}{z\sqrt{8\pi\sigma_{h_1}^2}}\exp\left[-\frac{\left(\xi\ln\left(z\right) - 2\mu_{h_1}\right)^2}{8\sigma_{h_1}^2}\right] \tag{35}$$

and

$$F_Y\left(\frac{\Upsilon R}{z}\right) = \frac{1}{2}\left(1 + \mathrm{erf}\left[\frac{\xi\ln\left(\frac{\Upsilon R}{z}\right) - 2\mu_{h_2}}{2\sqrt{2}\sigma_{h_2}}\right]\right). \tag{36}$$

Now, substituting (30) and (34) into (29) produces the ergodic outage probability of the PSR system over the log-normal channel, expressed as

$$\begin{aligned}
\mathcal{O}_{PSR}\left(C_{th}\right) = &1 - \frac{1}{2}\,\mathrm{erfc}\left[\frac{\xi\ln\left(\Xi R\right) - 2\mu_{h_2}}{2\sqrt{2}\sigma_{h_2}}\right] \\
&+ \frac{\xi}{\sqrt{32\pi\sigma_{h_1}^2}}\int\limits_{\Xi R}^{\infty}\exp\left[-\frac{\left(\xi\ln\left(z\right) - 2\mu_{h_1}\right)^2}{8\sigma_{h_1}^2}\right] \\
&\times\frac{1}{z}\left(1 + \mathrm{erf}\left[\frac{\xi\ln\left(\frac{\Upsilon R}{z}\right) - 2\mu_{h_2}}{2\sqrt{2}\sigma_{h_2}}\right]\right)\mathrm{d}z. \tag{37}
\end{aligned}$$

## V. Ergodic Outage Probability Analysis of the Ideal Relaying Receiver System

Unlike the TSR and PSR protocols, the IRR scheme is capable of, concurrently, processing information and harvesting energy from the same received signal. Therefore, the signal received at the information receiver of the relay is expressed as

$$y_r(t) = \sqrt{\frac{P_s}{d_1^m}} h_1 s(t) + n_r(t),\tag{38}$$

and the harvested energy and the relay transmit power can be given, respectively, as

$$E_H = \frac{\eta P_s h_1^2}{d_1^m}(T/2)\tag{39}$$

and

$$P_r = \frac{2E_H}{T} = \frac{\eta P_s h_1^2}{d_1^m}.\tag{40}$$

Using (40), the received signal at the destination can be written as

$$y_d(t) = \sqrt{\frac{\eta P_s}{d_1^m d_2^m}} h_1 h_2 \,\bar{s}(t) + n_d(t).\tag{41}$$

Now, using (38) and (41), the SNRs at the relay and destination nodes in the IRR system can be respectively expressed as

$$\gamma_r = \frac{P_s h_1^2}{d_1^m \sigma_r^2}\tag{42}$$

and

$$\gamma_d = \frac{\eta P_s h_1^2 h_2^2}{d_1^m d_2^m \sigma_d^2}.\tag{43}$$

The derivation of the ergodic outage probability of the IRR system is omitted in this paper due to space limitation. However, using (42) and (43), and following same procedure as in the previous section, it is straightforward to show that

$$\begin{aligned}
\mathcal{O}_{IRR}(C_{th}) =& 1 - \frac{1}{2}\operatorname{erfc}\left[\frac{\xi\ln(\Phi R) - 2\mu_{h_2}}{2\sqrt{2}\sigma_{h_2}}\right]\\
&+ \frac{\xi}{\sqrt{32\pi\sigma_{h_1}^2}}\int_{\Phi R}^{\infty}\exp\left[-\frac{(\xi\ln(z) - 2\mu_{h_1})^2}{8\sigma_{h_1}^2}\right]\\
&\times\frac{1}{z}\left(1 + \operatorname{erf}\left[\frac{\xi\ln\left(\frac{\Psi R}{\Theta z}\right) - 2\mu_{h_2}}{2\sqrt{2}\sigma_{h_2}}\right]\right)dz.\tag{44}
\end{aligned}$$

where $R = 2^{2C_{th}} - 1$, $\Phi = P_s/\left(d_1^m \sigma_r^2\right)$, $\Theta = \eta P_s$ and $\Psi = d_1^m d_2^m \sigma_d^2$.

## VI. NUMERICAL RESULTS

This section presents some numerical examples of the derived expressions along with Monte Carlo simulations. The system parameters adopted here, unless clearly stated otherwise, are as follows: $P_s = 1$ watt, $\sigma_i^2 = 3$ dB, $\mu_i = 3$ dB ($i\epsilon\{1,2\}$), $\sigma_d^2 = 2\sigma_{a,r}^2 = 2\sigma_{c,r}^2 = 0.01$ watt, $\eta = 1$ and $m = 2$.

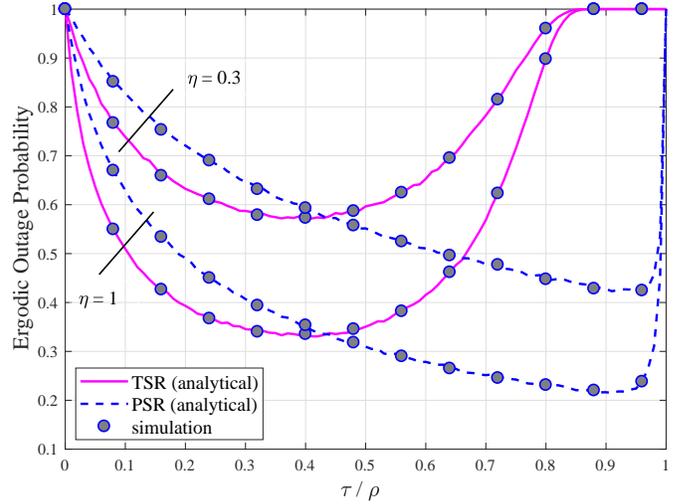

Figure 4: Ergodic outage probability performance versus the energy-harvesting time and power-splitting factors for the TSR- and PSR-based DF relaying systems over the log-normal fading channel.

### A. The Impact of $\tau$ and $\rho$ on the Ergodic Outage Probability

This section investigates the effect of the energy-harvesting time and the power-splitting factors on the ergodic outage probability of the TSR- and PSR-based systems. Fig. 4 shows some numerical examples and simulated results of the ergodic outage probability for the TSR and PSR systems as a function of the energy-harvesting time and power-splitting factors with different values of $\eta$ when $d_1 = d_2 = 5$ m. The analytical results of the TSR and PSR systems are obtained from (19) and (37), respectively. The first observation once can clearly see from these results is that increasing the energy-harvester efficiency will always enhance the performance regardless of the energy-harvesting protocol deployed. It is also apparent that when the energy-harvesting time or power-splitting factor approaches either zero or one, the performance degrades significantly due to the fact that the harvested power it either too small or unnecessarily too large. Therefore, a good selection of these parameters is crucial to achieve best performance.

### B. Performance Optimization

In this section, we present results for the optimized TSR, optimized PSR and IRR-based schemes. Although the optimal energy-harvesting time factor, $\tau^*$, and the optimal power-splitting factor, $\rho^*$, that minimize the ergodic outage probability cannot be easily expressed in closed-form, it is straightforward to find the solution numerically from (19) and (37) using some software tools. Substituting the optimal values of $\tau^*$ and $\rho^*$ in (19) and (37), respectively, and varying the threshold value, we obtain the optimal ergodic outage probability of the TSR and PSR approaches, as illustrated in Fig. 5. In addition, the ergodic outage probability for the IRR system, obtained from (44), are also included on this plot. This figure shows the ergodic outage probability for two different source-destination distances when $\eta = 1$. It should be pointed out that for all the considered distances in this section, the relay is placed at the midway between the source and destination nodes. It can be seen that the IRR system always outperforms the TSR and PSR schemes, and that the optimized PSR system has always better performance than that of the TSR approach.

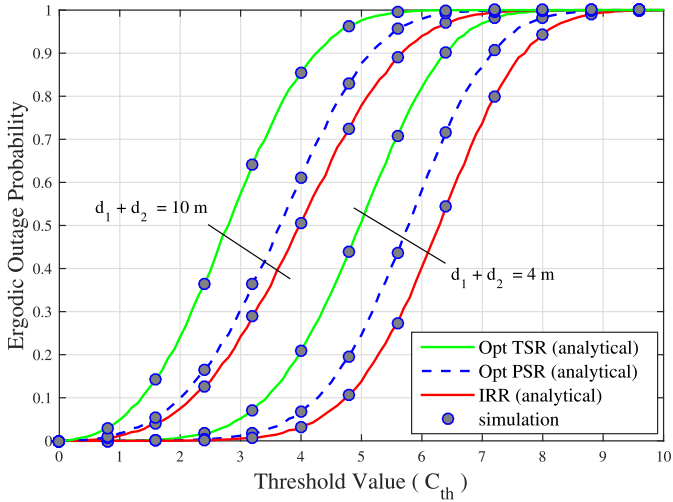

Figure 5: Optimal ergodic outage probability performance versus the threshold value for the optimized TSR, PSR and IRR systems with various source-destination distances.

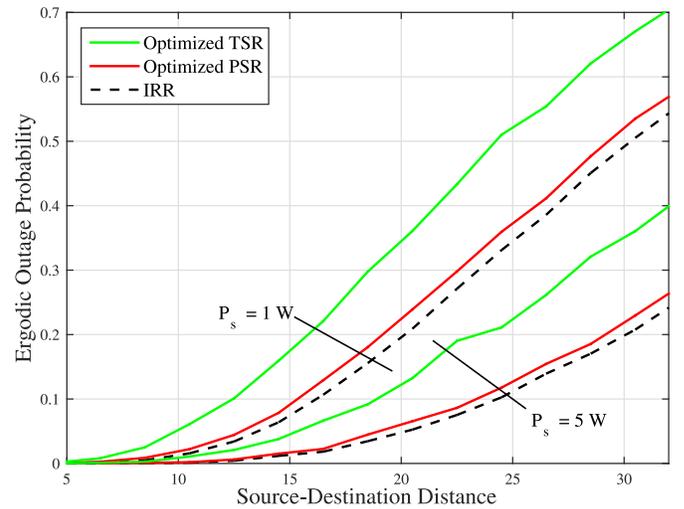

Figure 6: Optimal ergodic outage probability performance with respect to the source-destination distance for the TSR, PSR and IRR systems with different values of the source transmit power.

Furthermore, Fig. 6 depicts the optimal outage probability for the three systems with respect to the source-destination distance for different values of the source transmit power. Clearly, as the source transmit power is increased, the performance improves for all the considered systems. It can also be noticed that as the nodes become further apart, the probability performance deteriorates.

## VII. CONCLUSION

This paper analyzed the performance of energy-constrained dual-hop relaying networks over log-normal fading channels. DF relaying was deployed and three energy-harvesting protocols were studied, namely TSR, PSR and IRR. The system performance is evaluated in terms of the ergodic outage probability for which analytical expressions were derived. The good agreement between the numerical examples and the Monte Carlo simulations clearly confirm the accuracy of our analysis. Results showed that optimizing the energy-harvesting time (in the TSR protocol) and the power-splitting factor (in the PSR protocol) is the key to achieve best performance. In addition, the optimized PSR system has shown to have close performance to that of the IRR approach. It was also demonstrated that increasing the transmit source power and/or the energy-harvester efficiency can positively impact the system performance.